\documentclass[conference]{IEEEtran}
\IEEEoverridecommandlockouts
\usepackage{cite}
\usepackage{amsmath,amssymb,amsfonts}
\usepackage{algorithmic}
\usepackage{graphicx}
\usepackage{textcomp}
\usepackage{xcolor}
\newcommand{\subparagraph}{}

\usepackage{titlesec}

\def\BibTeX{{\rm B\kern-.05em{\sc i\kern-.025em b}\kern-.08em
    T\kern-.1667em\lower.7ex\hbox{E}\kern-.125emX}}

\newcommand{\shrink}{Eliminate vertical white-space}
\newcommand{\vshrink}[1]{
	\ifdefined\shrink 
	\vspace{-#1cm}
	\else
	\vspace{0cm}
	\fi
}

\titlespacing{\section}{0pt}{\parskip}{-\parskip}

\begin{document}

\title{Holistic Hardware Security Assessment Framework: A Microarchitectural Perspective}

\author{Tochukwu Idika \qquad Ismail Akturk \\
\textit{Electrical Engineering and Computer Science},
\textit{University of Missouri-Columbia}\\
tnifgf@mail.missouri.edu \qquad akturki@missouri.edu}

\IEEEaftertitletext{\vspace{-1\baselineskip}}

\maketitle

\begin{abstract}

Our goal is to enable holistic hardware security evaluation from the microarchitectural point of view. To achieve this, we propose a framework that categorizes threat models based on the microarchitectural components being targeted, and provides a generic security metric that can be used to assess the vulnerability of components, as well as the system as a whole.

\end{abstract}

\vshrink{0.2}

\section{Introduction}

For decades, security aspect of systems has been overlooked behind the unceasing effort to improve
the performance. As performance becomes restricted by the degree of energy-efficiency, the power management optimizations (e.g., dynamic voltage and frequency scaling) start to play a significant role in microarchitectural design and constraints. However, these optimizations may create side channels. Differential and leakage power analysis are the two types power monitoring side channel attack~\cite{baddam}. Differential power analysis (DPA) utilizes the correlation between input data and the dynamic power consumption of the microarchitectural components~\cite{avirneni}, whereas the leakage power analysis (LPA) employs the correlation between the input data and leakage power dissipation of the given component~\cite{alioto2010}. 
Evidently, hardware itself becomes a target for security attacks (especially by exploiting the side effect of energy/power optimizations in hardware) forcing system architects to consider the
security as first-class design constraint, similar to performance and energy efficiency. While such realization is promising, the existing threat model descriptions are ad-hoc, and security metrics are ill-defined (or non-existing), making it challenging to have fair and consistent vulnerability analysis among different microarchitectural components, optimizations, and system architectures. 
To address this challenge, we propose a framework that provides systematic categorization of threat models, and describes a security metric formulation which can be used to evaluate vulnerability of microarchitectural components, as well as the system as a whole.


For many microarchitectural exploration and assessment, we are used to rely on well-defined set of applications, such SPEC Benchmarks~\cite{spec2006}. These benchmarks allow architects to assess the impact of different microarchitectural components (and system as a whole) under various execution patterns on performance and/or energy-efficiency. Each benchmark can stress a particular microarchitectural component, and the evaluation can be reported in a well-defined, standardized way that allows fair comparison of alternatives. However, currently, we have no equivalent mechanism to perform hardware security assessment, neither at microarchitectural-level, nor at the system-level. The proposed framework would fill this gap. In this framework, there will be standardized definition of threat models (similar to benchmarks), and unified (i.e., universal) security metric that allows comparison of vulnerabilities of microarchitectural components (and whole system) to the respective threat models. In the following sections, we detail how threat models can be classified from a microarchitectural perspective, and how a security metric can be developed to be used as an individual (and global) assessment tool. 


\section{Classification of Threat Models}
Each threat model should be described in a standardized way that would allow system architects to reproduce the attack on the system being evaluated. Similar to applications in benchmark suites, a particular threat model may target different microarchitectural component (e.g., cache, branch predictor, memory). So, an architect who is interested in assessing a particular component's security vulnerability should be able to identify which threat models are targeting that particular component. 

\begin{figure}[h]
	\centering
	\includegraphics[width=1\columnwidth]{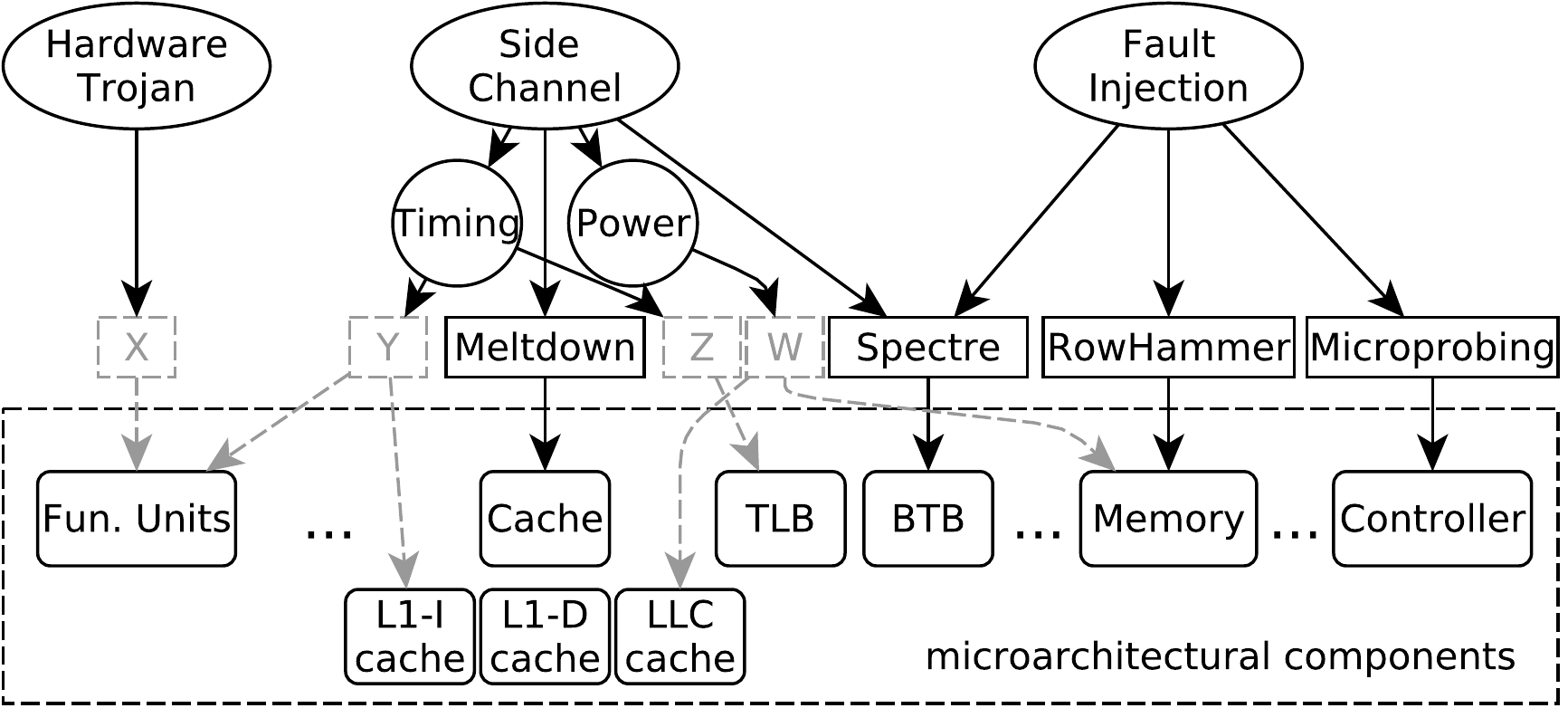}
	\vshrink{0.6}
	\caption{Microarchitectural-based threat model categorization. X, Y, Z, W should be known threat models, for illustration purposes, here they are shown generically.}	
	\label{fig:classification}
\end{figure}

In the proposed framework, each threat model can be categorized with respect to the microarchitectural component it targets. Fig.~\ref{fig:classification} illustrates such a categorization. As an example, if an architect is trying to assess the security vulnerability of the cache in his/her design, then (s)he has to experiment with threat models Y (a timing side channel attack), meltdown (side channel attack) and W (a power side channel attack). Without experimenting with all known attacks targeting the given component, it is impossible to make a fair and consistent assessment with respect to other alternative design choices for that particular microarchitectural component.

\vshrink{0.2}
\vshrink{0.2}

By using such a categorization given above, it is also possible to have holistic security assessment of a given system. In its simplest form, it would be an aggregation of evaluations performed for each microarchitectural component in the given system. 
To make a quantitative assessment and aggregation of separate assessments (for each component), we need a unified security metric that is representative for all possible components that can exist in any system and all threat models defined in the framework. In the following, we discuss how such unified security metric can be defined and used for holistic hardware security assessment.

\section{Security Metric for Holistic Assessment}

The previous studies proposed separate (and mostly incompatible) metrics for different attacks targeting different components~\cite{primer2014,rostami2013}, hence, there is no comprehensive way of combining the results to have holistic hardware security assessment of an entire system. To address that, we propose a quantitative metric that 
abstracts software and network layers of the system and focuses the vulnerability of microarchitectural components.
The proposed metric is generic, but universal, and called Vulnerability Factor (VF). 
It is similar in essence to side-channel vulnerability factor (SVF)~\cite{scv2012} that quantifies patterns observed by the attackers’ and measures the correlation to the victim’s actual execution patterns. Cache side channel vulnerability metric (CSV)~\cite{zhang2013} is a narrow but a more pragmatic version of SVF. Despite the similarities, proposed VF it is not specific to a particular attack model (as opposed to~\cite{scv2012} and~\cite{zhang2013}). In general, VF describes a metric that allows to measure the correlation between an operation performed on a particular microarchitectural component and the observation made by a specific threat model targeting the given component. Each component in the system has a VF under a given threat model. So, the overall VF for a component would become the aggregation of separate VFs under different threat models. Assume a set of components ($Component Set$) and a set of threats ($Threat Set$):
\vshrink{0.3}
$$VF_{comp_i} = \sum_{t=1}^{N}VF_{i,t} \times w_{t} $$
\noindent where $VF_{comp_i}$ represents the aggregated vulnerability factor of \textit{i'th component} ($comp_i \in Component Set$), $VF_{i,t}$ represents the vulnerability factor of \textit{component i} under the \textit{threat model t} (\textit{threat model t} $\in Threat Set$ and $1\leq t \leq N$, where N is the number of threat models known), and $w_{t}$ represents a weight of the given \textit{threat model t}. The weights exist to provide flexibility to an architect to set relative importance for the threats -- if all threats are equally important for an architect, then the weights will be same. When reporting the $VF_{comp_i}$, it is expected to report associated weights as well (default is 1).

The assessment of $VF_{comp_i}$ of different systems would allow an architect to evaluate how various design and implementation choices of the given component could impact its overall vulnerability.

Similarly, the VF of the whole system can be measured as the aggregation of VFs of each component exist in the system:
\vshrink{0.3}
$$HVF_x = \sum_{i=1}^{M}VF_{comp_i} \times w_{i} $$
\noindent where $AVF_x$ represents the holistic vulnerability factor of a given system (e.g., \textit{System X}), $VF_{comp_i}$ represents the aggregated vulnerability factor of \textit{i'th component} ($comp_i \in Compoonet Set_X$ and $1\leq i \leq M$, where M is the number of components in the system), and $w_{i}$ represents a weight of the given \textit{component i}. The weights have similar semantic as mentioned above.

Note that the actual derivation of VF may vary as component and/or threat model changes. Here, we do not advocate a single VF derivation, but we provide a generic metric whose specific derivation should be determined by the community. Once a particular VF for a component under a threat model is determined, it can be standardized as part of the proposed framework which would allow the holistic security assessment of a system from the microarchitectural perspective.

\section{Challenges and Future Work}


There are two fundamental challenges for quantitative holistic hardware security assessment.
The first challenge is the fact that security is a moving target, meaning that vulnerability assessments and quantification are limited by the threat models that we know today (i.e., there is no way to know all possible threats beforehand that may exist). As a result, a system deemed secure considering threat models that we know today, could prove to be totally insecure tomorrow when a new threat model is identified. 
When a new threat model is identified, it should be added into the proposed framework (similar to adding a new application into a benchmark suite). That would translate into a new version of the framework. As a result, an architect should also indicate which version of the framework (s)he used when reporting the vulnerability assessment of a system. We plan to maintain a timestamped database that keeps track of the known threat models that allows to distinguish different versions of the framework. 

The second challenge is the determination of vulnerability factor (VF) for each microarchitectural component and a threat model. Conceptually, VF is a metric that allows to measure the correlation between an operation performed on a particular microarchitectural component and the observation made by a specific threat model targeting the given component. However, a specific implementation may be needed to calculate the correlation on each component and a threat model. Identification of specific VFs left as a future work.

\bibliographystyle{ieeetr}
\bibliography{strings,references}

\end{document}